\documentclass{article}
\usepackage{spconf,amsmath,epsfig}
\usepackage{algorithm2e}

\DeclareMathOperator*{\argmax}{arg\,max}

\title{Collaborative Training in Sensor Networks:
\\A graphical model approach}
 \name{Haipeng Zheng, Sanjeev R. Kulkarni and H. Vincent Poor}
 \address{Department of Electrical Engineering, Princeton University\\
 Princeton, NJ, 08544}
\begin{document}
%\ninept
%
\maketitle
\begin{abstract}
Graphical models have been widely applied in solving distributed inference
problems in sensor networks. In this paper, the problem of coordinating
a network of sensors to train a unique ensemble estimator under
communication constraints is discussed.  The information structure of
graphical models with specific potential functions is employed, and this
thus converts the collaborative training task into a problem of local
training plus global inference. Two important classes of  algorithms of graphical model inference, message-passing algorithm and sampling algorithm, are employed to tackle low-dimensional, parametrized and high-dimensional, non-parametrized problems respectively. The efficacy of this approach is demonstrated by concrete examples.
\end{abstract}
\section{Introduction}
\label{sec:intro}
It is widely recognized that distributed inference methods developed for
graphical models comprise a principled approach for information fusion in
sensor networks (see \cite{cetin2006}). With powerful graphical
model (GM) inference tools at hands, the similarity
between sensor networks and graphical models compels researchers to
model sensor network problems using graphical
models so that the deeply researched GM algorithms, including
sum-product, sampling and variational algorithms, can be applied
to sensor networks.

Although this analogy seems to be simple, the map from sensor
network problems to graphical models is not straightforward. As
pointed out in \cite{cetin2006}, it is the \emph{informational
structure} of the distributed inference problem, involving the
relationships between sensed information and the variables about
which we wish to perform estimation, that is just as critical as the
communication structure of the problem. How to model distributed
inference problems as graphical models is as important as solving
the problem itself. A wide range of distributed inference problems
have been reformulated with graphical models, including
self-localization \cite{Ihler2005a}, multi-object data
association and tracking \cite{Chen2006}, \cite{Uney2007},
distributed hypothesis testing \cite{null2006}, and nonlinear
distributed estimation \cite{Chong2004}.

One of the advantages of modeling distributed inference problems in
sensor networks as inference problems on graphical models is to find
communication-efficient ``messages" that are exchanged among the
sensors. In many ad-hoc algorithms for distributed inference
problems, the messages transmitted among the sensors are
problem-specific. If we can successfully model these problems as
graphical models, the messages exchanged among the sensors turn out
to be exactly the messages specified by the corresponding graphical
model message-passing algorithms such as the sum-product algorithm.

Another issue, much more important in the area of wireless sensor
networks than in graphical models, is the communication cost. In
wireless sensor networks, communication is usually constrained and
expensive, unlike in centralized inference algorithms, where
message-passing is almost free. This difference leads to totally
different optimization objectives in these two areas, even for
exactly the same graphical models. In sensor network problems, we
look for inference algorithms with more local computation and less
message-passing; in graphical models, we are interested in
algorithms that minimize the computational complexity. This brings
us new problems like how to decrease the amount of data exchanged
among the sensors, as described in \cite{cetin2006}, \cite{Kreidl2006},
 \cite{Ihler2005}, and \cite{Sudderth2003}. This
problem is unique for graphical models applied to distributed
inference problems.

There are many distributed inference problems that have not been
described in the language of graphical models. One type of such
problem falls into the category of distributed
learning/collaborative training, as described by Predd, et al., in
\cite{Predd2009}, \cite{Predd2006a}, and \cite{Predd2006b}. In these
problems, the sensors collaboratively train their individual
estimators so as to minimize the training error of a kernel
regression, subject to consistency of prediction on shared data. In
\cite{Predd2006a}, an iterative algorithm is designed to achieve
this training goal.

In our paper, we aim to model the distributed training problem
as an inference problem on graphical models. In these settings, independent and identically distributed (i.i.d.) data are collected by different sensors, which are able to communicate with each other under some constrains. The sensors, without directly sharing their training data (usually high-dimensional, confidential and in large amount), attempt to collaboratively find a good ensemble classifier/estimator. In our framework, we manage to transform collaborative training problem, usually solved by an ad-hoc design (e.g. alternating projection) in a sensor network, into an
\emph{inference problem} on a graphical model \emph{combined with
local training}. This conversion from ad-hoc collaborative training
to local training plus collaborative inference is due to the
application of the graphical model on a functional space of classifiers/estimators. The problem of selecting the optimal estimator is converted into a maximum a posteriori probability (MAP) problem on a graphical model, with each random variable supported on a functional space to which
the estimators belong.

\section{From collaborative training to graphical models}
The self-localization  problem in \cite{cetin2006} provides us with an excellent example of how to convert a distributed inference problem in a sensor
network into an inference problem in a graphical model. Similar to
this scheme, we design our graphical
model for the collaborative training task to convert a ``global
collaborative training" into a ``local training plus global
inference" problem.

\begin{figure}[h]
\begin{center}\leavevmode
\includegraphics[width = 1.3 in]{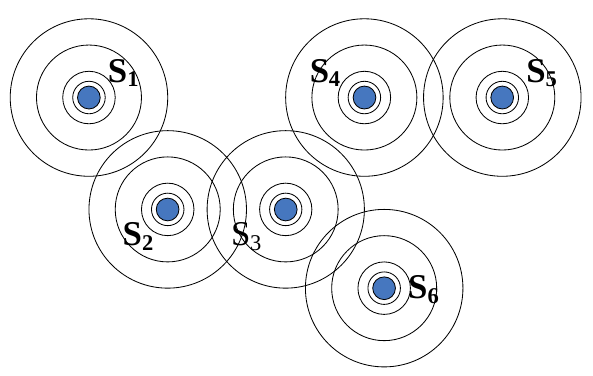}
\includegraphics[width = 1.3 in]{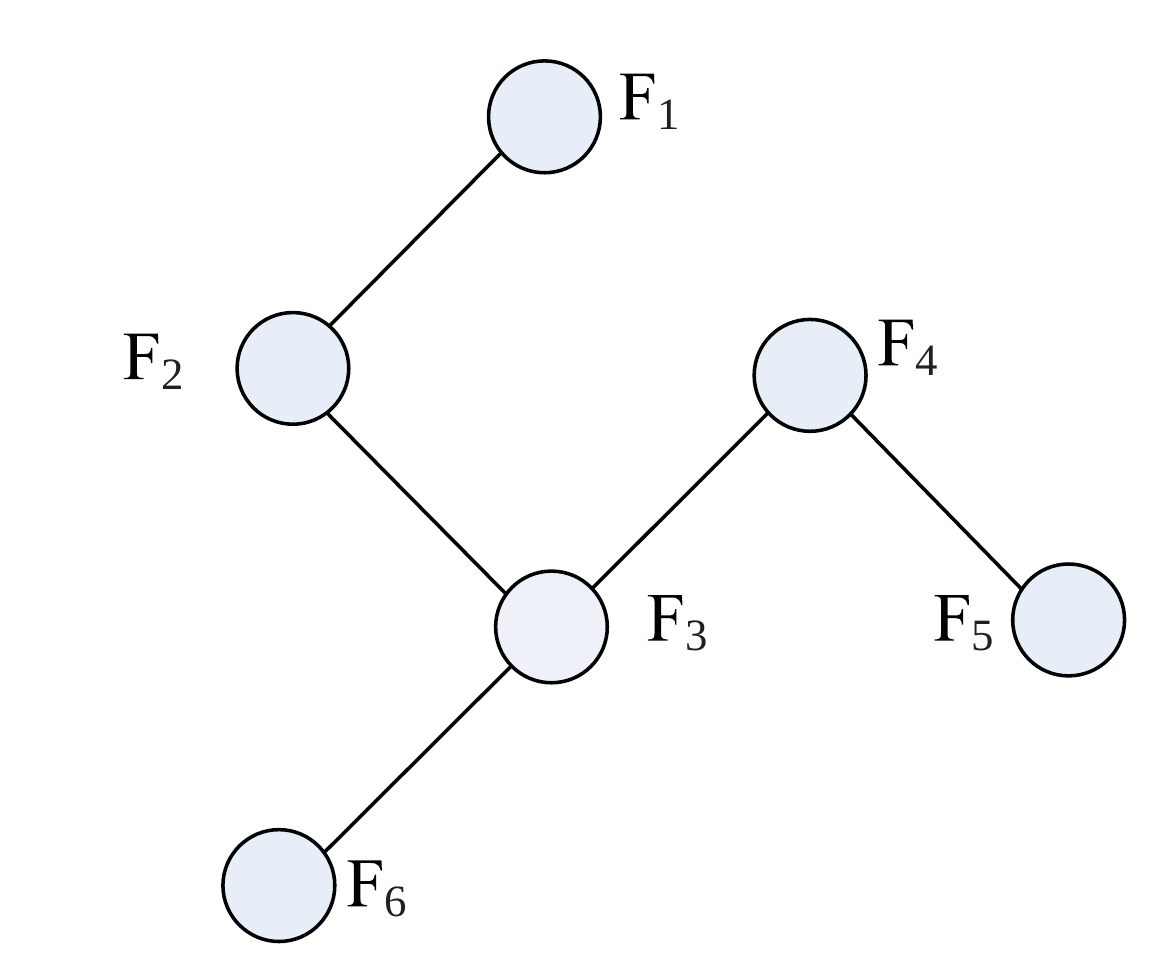}
\end{center}
\caption{A typical sensor network and its corresponding graphical
model abstraction.}\label{fig1}
\end{figure}

To make our model more concrete, we first illustrate the system in
Fig. \ref{fig1}, where there are 6 sensors, with limited
communication capability.

In Fig. \ref{fig1}, we abstract the sensor network into a graph. The
edges among the nodes represent the condition that the nodes are
able to communicate with each other. Now, we assume that each sensor
$s$ maintains a distribution of estimators, i.e., a random variable
$F_s$ supported by functional space $\mathcal{H}$, with
probability distribution $\rho_s(f_s)$.

Then, we assign a potential $\sigma_{s,t}(f_s,f_t)$ to the edge
between two connected sensors $s$ and $t$. Here,  $\sigma_{s,t}$ is
a requirement of the similarity between the estimators maintained by
adjacent sensors.

Based on these assumptions, the potential of the entire graphical
model is of the form
\begin{equation}
p(f) =
\frac{1}{Z}\prod_{s\in\mathcal{V}}\rho_s(f_s)\prod_{(s,t)\in\mathcal{E}}\sigma_{s,t}(f_s,f_t).
\label{eqn}
\end{equation}
If the graph is loopless, a potential of the form (\ref{eqn}) is a
standard form to apply message-passing algorithms, such as the sum-product algorithm. These algorithms enable us to find the marginal distribution/MAP of
the estimator at each sensor in a distributed way.

However, it is quite common that loops exist in the sensor network,
yet it is very difficult to do triangulation in a system without
centralized computation. To allow the use of
message-passing algorithms, Willsky, et al. apply \emph{loopy
belief propagation} described in \cite{Pearl1988} and
\cite{Murphy1999}. Although Jordan and Murphy in \cite{Murphy1999}
have shown some cases where loopy belief propagation might lead to
erroneous results, Willsky et al. have proven that under certain
conditions, loopy belief propagation is a contractive map (for some
specially defined distances) and hence converges to a unique limit.
Therefore, loopy belief propagation can be applied to make inferences
on the graphical model with potential (\ref{eqn}). Moreover, sampling algorithms are not affected by loops, and can always be applied readily.

Under our framework, the messages or samples passed among the sensors are no
longer data instances, but distributions or individual samples of functions. Thus, the sensors are no longer working with individual data points, but summaries of data - trained classifiers/estimators. Moreover, the message-passing algorithm ends
in finite steps for loopless graphs, which requires no iteration.
These advantages are due to the introduction of the graphical model.

\section{Local Training and Global Inference}
Based on the model described in the previous section, the problem
now can be separated into two stages:
\begin{enumerate}
\item Local training: find reasonable potentials $\rho_s$ and
$\sigma_{s,t}$;
\item Global inference: based on the potentials, compute the marginal
distribution of estimators at each sensor.
\end{enumerate}
We now discuss the details of these two stages.
\subsection{Local Training for Potentials}
There are two ways to obtain the potential $\rho_s$ based on local
data. If there is a good model or prior knowledge of the parameters
of $f_s$, then we can use local data to find $\rho_s$. However, in
most of the cases, the distribution of parameters to be estimated is
rather hard to compute explicitly. In this case, we can simply
employ a bootstrap algorithm to re-sample the local training data
for each individual sensor, and locally train a group of estimators,
which can be used to approximate the distribution of the estimator.
Thus, $\rho_s$ can be specified in a parametric (using the sample
estimators to estimate the distribution of parameters) or
non-parametric way (using several typical estimators as
``particles").

Finding $\sigma_{s,t}$ is more complicated. This is because in many
cases, $\sigma_{s,t}$ is related to the statistical properties of the
global estimator to be determined - i.e. it might be hard to
estimate locally. In this paper, we restrict our discussion to problems where the entire system has one unique  hypothesis. Thus, $\sigma_{s,t}$ is simply predetermined as
\begin{equation}
\sigma_{s,t}(f_s,f_t)=\left\{
\begin{array}{ll}
  1 & \mbox{$f_s$ and $f_t$ are the same} \\
  0 & \mbox{otherwise} 
\end{array}
\right.
\label{dirac}
\end{equation}
although in practice, we sometimes need to relax the peaky function slightly. With the assumption above, we can discuss two classes of algorithms in graphical models for our collaborative training, message-passing (for parametrized, low-dimensional cases) and sampling (for non-parametrized, high-dimensionally cases).

Interestingly, it can be shown that for a simple hypothesis testing problem ($H_0$ vs. $H_1$), with the similarity function defined as in (\ref{dirac}), the likelihood ratio methods (using the entire data set) and the collaborative training methods (computing the likelihood based on local data, and find marginal distribution by global inference) result in exactly the same outcome, given that the data collected by the sensors are i.i.d. To some extent, this supports the choice of our similarity function and the optimality of the scheme.

\subsection{Message-Passing for Parametrized Cases}
Message-passing is an accurate inference algorithm on graphical models. Here we assume that  the classifier/estimator of sensor $s$ can be parametrized by parameter $x_s$.

Then, according to the sum-product algorithm described in
\cite{Wainwright2005}, given the potential (\ref{eqn}), the
marginal at node $s$ is of the form
\begin{equation}
p(x_s)\propto\rho(x_s)\prod_{t\in\mathcal{N}(s)}M_{ts}^*(x_s),
\end{equation}
where $M_{ts}^*(x_s)$ is the message from sensor $t$ to sensor $s$.
When the graph is loopless, the sum-product algorithm
will converge in finitely many steps and converge to a unique limit.
However, for arbitrary graph structure, the sum-product form might only
be an approximation. Here we directly apply the message recursion
update formula for the sum-product algorithm:
\begin{equation}
M_{ts}(x_s)\leftarrow \sum_{x'_t}\left\{ \sigma_{s,t}
(x_s,x'_t)\rho(x'_t)\prod_{u\in\mathcal{N}(t)\backslash
s}M_{ut}(x'_t)
 \right\}.
\label{msgUpdate}
\end{equation}
Specially, if we assume that the potentials are of Gaussian form,
\begin{equation}
\rho_s(x_s) = \exp\left\{-\frac{(x_s-\mu_s)^2}{2\sigma_s^2}\right\}
\end{equation}
and
\begin{equation}
\sigma_{s,t}(x_s,x_t) =
\exp\left\{-\frac{(x_s-x_t)^2}{2\lambda_{s,t}^2}\right\},
\label{similarity}
\end{equation}
where $\lambda_{s,t}$ is close to $0$ to ensure consensus, then the message is also of a Gaussian form:
\begin{equation}
M_{ts}(x_s) = \exp\left\{-\frac{(x_s -
\mu_{ts})}{2\sigma_{ts}^2}\right\}.
\end{equation}
Therefore, message updating, as described in (\ref{msgUpdate}), can
be simplified to an update of the parameters $\mu_{ts}$ and
$\sigma_{ts}^2$, as specified below:
\begin{equation}
\mu_{ts} = \frac{\mu_t/\sigma_t^2+\sum_{u\in\mathcal{N}(t)\backslash
s}\mu_{ut}/\sigma_{ut}^2}{1/\sigma_t^2+\sum_{u\in\mathcal{N}(t)\backslash
s}1/\sigma_{ut}^2},
\end{equation}
and
\begin{equation}
\sigma_{ts}^2 = \lambda_{t,s}^2+\left(  1/\sigma_t^2 +
\sum_{u\in\mathcal{N}(t)\backslash s}1/\sigma_{ut}^2  \right)^{-1}.
\end{equation}

After the messages converge, the marginal distribution $p_s(x_s)$ of
each sensor $s$ is still a Gaussian distribution with
parameters given by
\begin{equation}
\hat{\mu}_{t} =
\frac{\mu_t/\sigma_t^2+\sum_{u\in\mathcal{N}(t)}\mu_{ut}/\sigma_{ut}^2}{1/\sigma_t^2+\sum_{u\in\mathcal{N}(t)}1/\sigma_{ut}^2},
\end{equation}
and
\begin{equation}
\hat{\sigma}_{t}^2 = \left(  1/\sigma_t^2 +
\sum_{u\in\mathcal{N}(t)}1/\sigma_{ut}^2 \right)^{-1}.
\end{equation}

It can be shown that when $\lambda_{s,t}\rightarrow 0$, and the
network is loopless, the MAP estimation of each sensor after running message-passing converges to the average
of $\mu_s$ weighted by $1/\sigma_s^2$, which is exactly how we combine i.i.d. Gaussian observations with different variances. In this sense, our scheme finds the optimal solution.

\subsection{Sampling Algorithms for High-dimensional Cases}
Message-passing for the parametrized case is straightforward. However, when the classifiers/estimators reside in some high-dimensional space (very common for most learning problems) and cannot easily be parametrized (like neural networks and decision trees), it is difficult to update the parameters directly and message-passing can be unimplementable. In this case, we resort to sampling methods to effectively search for the optimal classifiers/estimators.

The first problem we face is to find an expression for the distributions of estimators $f_s$ of individual sensors. Usually, by bootstrapping, we can obtain a group of ``particles'' $\{h_{sj}\}_{j=1}^{n_s}$ of the distribution of $f_s$. If we define a kernel $K(\cdot,\cdot)$ (non-negative), i.e. a measure of similarity of the estimators, then we can write the distribution of $f_s$ as
\begin{equation}
\rho_s(f_s) = \sum_{j=1}^{n_s}K(h_{sj},f_s).
\end{equation}
If we assume that the true hypothesis is unique, then we need to enforce consensus among the sensors; thus we define the similarity function as in (\ref{dirac}).

With these assumptions, the marginal distribution of any sensor $t$ in (\ref{eqn}) has the form (after summing out all the other variables)
\begin{equation}
p(f_t)=\prod_{s=1}^M\sum_{j=1}^{n_s}K(h_{sj},f_t).
\end{equation}
Therefore, the accurate MAP solution for the collaboratively trained estimator is given by
\begin{equation}
f^* = \argmax_{f\in\mathcal{H}_s}\prod_{s=1}^M\sum_{j=1}^{n_s}K(h_{sj},f),
\label{opt}
\end{equation}
where $M$ represents the total number of sensors and $n_s$ is the number of ``particles" bootstrapped by sensor $s$. Moreover, for simplicity, we define
\begin{equation}
\mathcal{H}_s = \left\{h_{sj}|s \in \{1,\ldots,M\}, j\in\{1,\ldots,n_s\}\right\}.
\end{equation}
It is difficult to apply an accurate inference algorithm to solve (\ref{opt}), because the closed-form messages involve an increasingly intricate product-sum of the kernels. Therefore, we resort to sampling methods to distributively tackle problem (\ref{opt}).

For simplicity, we only discuss the Gibbs Sampling case, which, in our model, reduces to the following algorithm: 
~\\
~\\
\restylealgo{boxed}
\begin{algorithm}[H]
\textbf{Repeat}
    \begin{enumerate}
    \item Randomly select sensor $s$;
    \item Fix the sample values of  its neighbors $\{f_t\}_{t\in\mathcal{N}_s}$;
    \item Conditioned on $\{f_t\}_{t\in\mathcal{N}_s}$, resample $f_s$ based on \\ the distribution of $f$\\
    $
    \frac{1}{Z}\left(\prod_{t\in \mathcal{N}_s}\sigma_{s,t}(f_t,f)\right)
    \left(\sum_{j=1}^{n_s}K(h_{sj},f)\right)
    $
    \end{enumerate} ~~~~~~~~~~where $f\in\{h_{sj}|1\le j \le n_s\}\cup\{f_t|t\in\mathcal{N}_s\}$;\\    
\textbf{End}
\caption{Sampling Algorithm\label{alg}}
\end{algorithm}
~\\
In the algorithm, the function $\sigma_{s,t}(\cdot,\cdot)$ (non-negative) is the similarity function (a very peaky function). However, in practice, to make the sampling algorithm non-trivial and more forgiving, $\sigma_{s,t}$ should be relaxed, allowing some discrepancy between its two inputs, and thus preventing the algorithm from falling into a trivial solution. Moreover, notice that $\mathcal{N}_s$ represents the set of neighbors of sensor $s$, and the restrictions of the space in which $f$ resides is due to the locality constraints.

In practice, the algorithm might converge rather slowly. In that case, we can change the random resampling (step 3) of the algorithm into a deterministic optimization, i.e. we can replace step 3 by
\begin{equation}
f_s = \argmax_{f\in \mathcal{H}_s}\left(\prod_{t\in \mathcal{N}_s}\sigma_{s,t}(f_t,f)\right)
    \left(\sum_{j=1}^{n_s}K(h_{sj},f)\right).
\end{equation}
This algorithm is prone to falling into a local minimum, yet converges faster. It is almost a greedy approach to finding the solution of (\ref{opt}).

There is one further subtle issue in implementing the sampling algorithm. Since the kernel is defined on various forms of classifiers/estimators, it might be hard sometimes, say, to define the similarity/distance between a decision tree and a neural network. If we define the prediction  of estimator $f_1$ on the training set to be a vector $\mathbf{f}_1$ and the prediction of estimator $f_2$ on the training set to be a vector $\mathbf{f}_2$, then we use the similarity between these two vectors as that of the two estimators.

In the collaborative training scenario, however, the entire training set is not accessible to each individual sensor; thus $K$ and $\sigma_{s,t}$ can only be estimated locally, i.e., sensor $s$ can only compute $K$ and $\sigma_{s,t}$ based on its own data. Therefore, in the above algorithm, the kernels are actually subscripted. We will show empirically that this local data restriction indeed compromises the performance of the system, yet this is the price we pay for distributed algorithms.

The sampling algorithm has the advantage that only a properly selected kernel (a measure of similarity among classifiers) is required, unlike the case of message-passing, where we usually expect a linear Euclidean space of parameters. Moreover, sampling is not affected by the loops in the undirected graph model, and the sensors can update their samples asynchronously, as long as the samples of their Markov blankets are fixed.

\section{Experiments}
\subsection{Message-Passing Algorithm: Linear Regression}
Assume that $m$ sensors are distributed in the domain $[0,1]^2$. The sensors
cooperate to estimate the slope $k$ of a straight line $z = kx$. The
sensor at location $(x,y)\in [0,1]^2$ observes a noisy version of
the value of $z$ at $x$, and the noise at point $(x,y)$ is additive
and has a Gaussian distribution of variance $\sigma^2\sin^2(2\pi
x)$. We also assume that each sensor can query the value of
observations of its neighbors within a radius $r$.

For this problem, each sensor is capable of estimating the global
model based on its own observation and those of its neighbors -
because the slope $k$ can be estimated well even if we observe only
a small part of the straight line. So for this consensus problem,
the key step is to find the potential/distribution of each
individual sensor. Bootstrapping is a suitable method in this case. Each
sensor $s$ simply bootstraps over its accessible data (the data of
itself and its neighbors) and uses the sample distribution to
approximate the potential $\rho_s$. For computational simplicity, we
parametrize these distributions as Gaussian (even though this is
not accurate) so that we can simply apply the parameter update
formula derived from the previous section.

\begin{figure}[htb]
\begin{center}\leavevmode
\includegraphics[scale = 0.40]{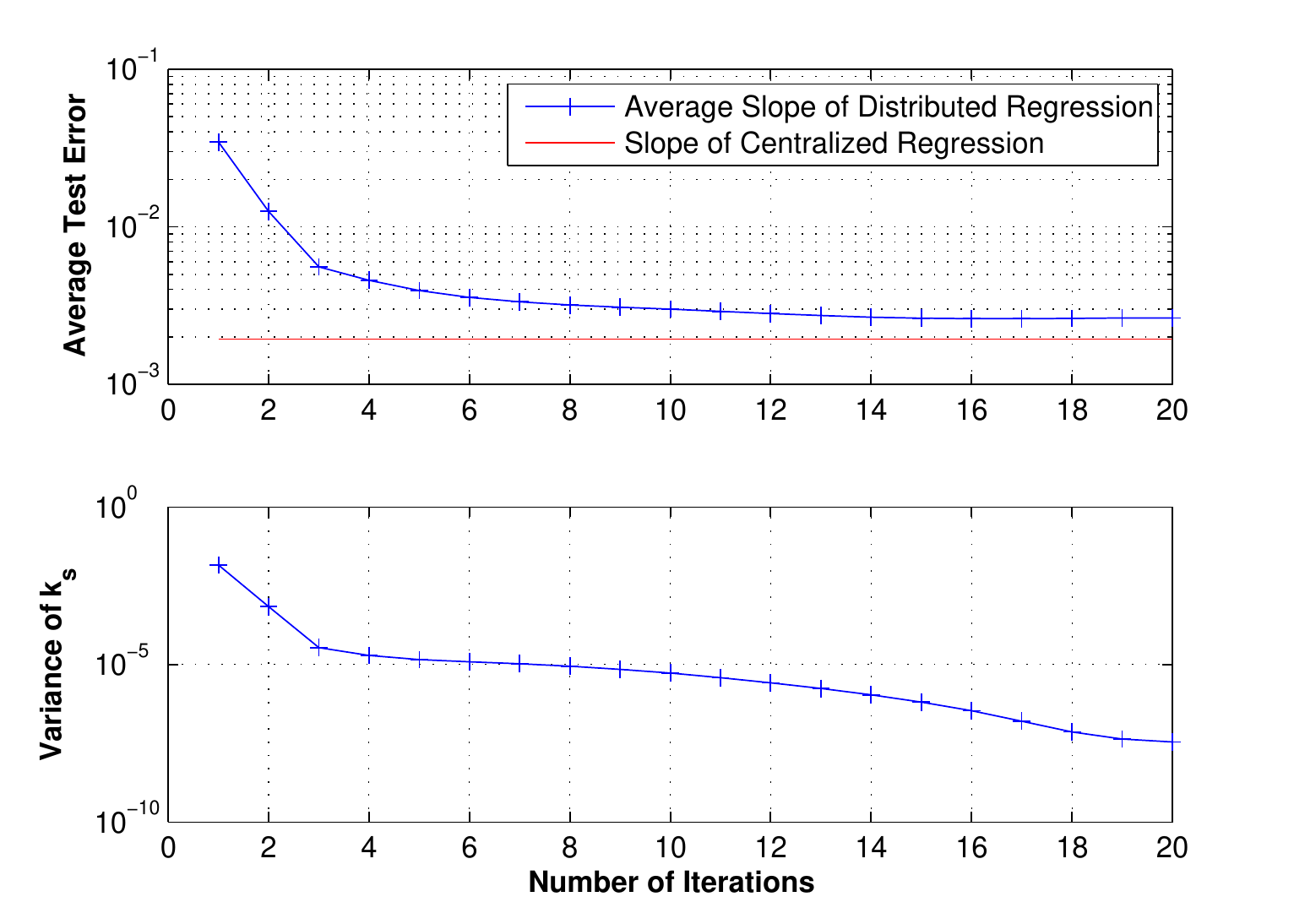}
\end{center}
\caption{Performance of the collaborative training algorithm running
on the sensor network. The plot on top depicts the decreasing
trend of the test error. The plot on the bottom demonstrates the
attenuation of the variance of the estimated parameter, indicating
the speed with which the system approaches consensus.}\label{cvg}
\end{figure}

There are 50 sensors with communication radius of 0.2 in this
consensus problem, i.e. $m = 50$, $r = 0.2$ and $\lambda_{s,t} = 0$.
A typical result of the simulation is shown in Fig. \ref{cvg}.

Note that the estimates of the slopes of different sensors in the
distributed system come to consensus rather quickly - the variance
of slopes among the sensors reduces to a negligible level after a
few rounds. On the other hand, the average test error decreases
quickly, very close to the performance of centralized linear
regression.

\subsection{Sampling Algorithm: Decision Tree Classifiers}

We select the Chess data set (King-Rook vs. King-Pawn), a 3196-instance, 36-dimension, 2-class data set from the UCI machine learning repository, for this experiment. We randomly select 2000 data points, evenly distributed at 20 different sensors, as the training set, and use the remaining 1196 data points as the test set. The communication topology of the 20 sensors is a random graph of expected degree of 4. And each sensor, by bootstrapping, generates 4 classifiers (chosen to be standard decision tree classifiers provided by MATLAB).
We define the kernel as
\begin{equation}
K(f_s,f_t)=\left(1-\frac{1}{n}\|\mathbf{f}_s-\mathbf{f}_t\|\right)^4,
\end{equation}
where $\mathbf{f}_s$ denotes the vector of prediction of classifier $f_s$ on all the local training data points, $\|\cdot\|$ denotes the Hamming distance of two 2-symbol vectors, and $n$ is the total number of local data. Moreover, we select $\sigma_{s,t}(\cdot,\cdot)$=$K(\cdot,\cdot)^3$ to make it a properly peaky function.
 
\begin{figure}[htb]
\begin{center}\leavevmode
\includegraphics[scale = 0.40]{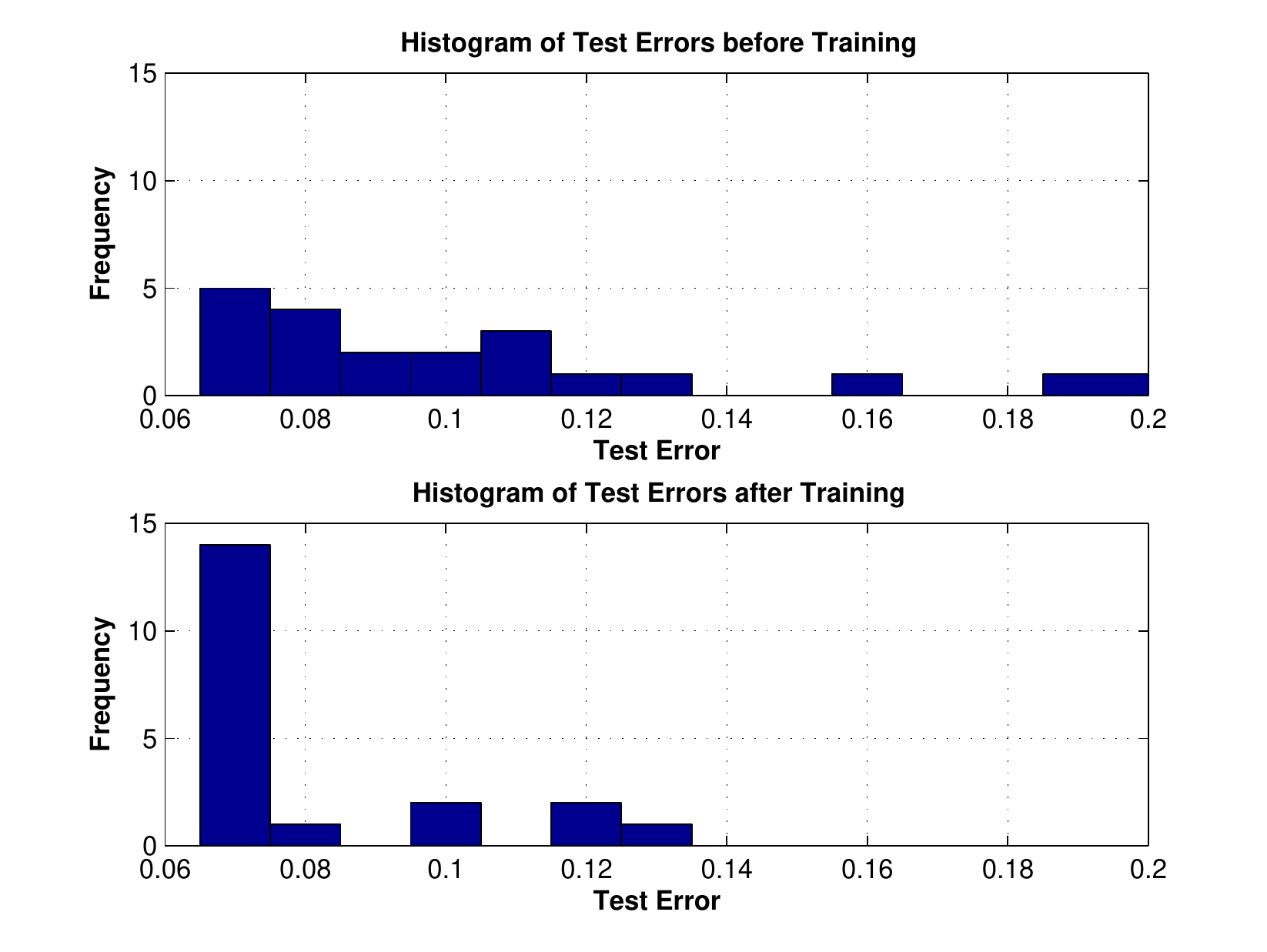}
\end{center}
\caption{Histograms of test errors among the sensors before and after running the sampling algorithm.}\label{hist}
\end{figure}

The sensors initialize their sampled classifiers by solving (\ref{opt}) based on their individual data (i.e., they solve the optimization problem of (\ref{opt}) without the product). Running the greedy version of the algorithm \ref{alg} for 4000 rounds, we obtain the results shown in Fig. \ref{hist} and Table \ref{tab1}.

\begin{table}[hbt]
\centering
\begin{tabular}{c |c | c  }

  \hline
  Data  & Algorithm & Test Error\\
  \hline
  Centralized &  Centralized decision tree & .0109\\
  Distributed & Centralized solution to (\ref{opt}) & .0702\\
  Distributed & Non-collaborative training  &  .0941 (median)\\
  Distributed & Sampling algorithm & .0702 (median)\\
  Distributed & Average of all classifiers & .0669\\
  \hline
\end{tabular}
  \caption{Test errors of different algorithms, compared with the result of the sampling algorithm.}
  \label{tab1}
\end{table}

As shown in the results, the sampling algorithm (based on Gibbs Sampling) enables a major portion of the sensors in the network to find the optimal classifier, with respect to the distributed data, centralized solution to (\ref{opt}), and much better than the results given by non-collaborative training. A simple average of all the bootstrapped classifiers (similar to bagging) seems to be slightly better, yet the generated classifier is much more complicated than the results of the sampling algorithm (80 trees vs. 1 tree). 

In this example, we have seen that our scheme of collaborative training can be quite effective even for a very complex, high-dimensional space of classifiers, without transmitting any training data points.

\section{Conclusions and Discussions}
We have applied our scheme to both  parametrized, low-dimensional cases and non-parametrized, high-dimensional cases, and accurate message-passing and approximate sampling algorithms demonstrate their efficacy for these two cases separately. Without directly sharing data, the sensors are able to reach consensus and collaboratively search for a classifier/estimator satisfying certain optimality properties.

Although the ``collaborative" part of our algorithms
is based on message-passing or sampling algorithms borrowed from graphical models, another essential step of our algorithm is local training, as we only briefly resort to bootstrapping in this paper.  It is of interest to find
more detailed statistical tools to estimate these potentials, or more specifically, the distributions of classifiers/estimators so that we
may be able to guarantee stronger optimality and obtain better
performance.

Despite the issues and challenges described above, we have shown the
efficacy of this framework with two different examples. It
is worthwhile to design more delicate algorithms and to prove stronger
results under this framework.

\bibliographystyle{IEEEbib}
\bibliography{GraphicalModels}

\end{document}